\documentclass[useAMS,usenatbib, final]{mn2e}
\usepackage[T1]{fontenc}
\usepackage[utf8]{inputenc}
\usepackage{aas_macros}
\usepackage{amsmath, amssymb}
\usepackage{booktabs}
\usepackage{graphicx}
\usepackage{txfonts}
\usepackage{xcolor}
\usepackage{xspace}
\usepackage{mathrsfs}
\usepackage{balance}
\usepackage{microtype}
\usepackage{hyperref}

\newcommand*{\IS}{$\mathrm{IS}$\xspace}
\newcommand*{\ISh}{$\mathrm{IS_h}$\xspace}
\newcommand*{\ISi}{$\mathrm{IS_i}$\xspace}
\newcommand*{\ISl}{$\mathrm{IS_l}$\xspace}
\newcommand*{\VD}{$\mathrm{VD}$\xspace}
\newcommand*{\VDh}{$\mathrm{VD_h}$\xspace}
\newcommand*{\VDi}{$\mathrm{VD_i}$\xspace}
\newcommand*{\VDl}{$\mathrm{VD_l}$\xspace}
\newcommand*{\CR}{$\mathrm{CR}$\xspace}
\newcommand*{\CRh}{$\mathrm{CR_h}$\xspace}
\newcommand*{\CRi}{$\mathrm{CR_i}$\xspace}
\newcommand*{\CRl}{$\mathrm{CR_l}$\xspace}
\newcommand*{\RD}{$\mathrm{RD}$\xspace}
\newcommand*{\RDh}{$\mathrm{RD_h}$\xspace}
\newcommand*{\RDi}{$\mathrm{RD_i}$\xspace}
\newcommand*{\RDl}{$\mathrm{RD_l}$\xspace}

\newcommand*{\Lx}{L_{\rm X}}
\newcommand*{\Lrot}{L_{\rm rot}}
\newcommand*{\Lstr}{L_{\rm v}}
\newcommand*{\Lsigma}{L_{\sigma}}
\newcommand*{\Lkin}{L_{\rm kin}}
\newcommand*{\Lsn}{L_{\rm SN}}
\newcommand*{\Lb}{L_{\rm B}}

\newcommand*{\emisx}{\varepsilon_{\rm X}}

\newcommand*{\mtot}{\dot\rho}
\newcommand*{\mSN}{\dot\rho_{\rm SN}}
\newcommand*{\mWinds}{\dot\rho_{\star}}

\newcommand*{\rh}{r_{\rm h}}

\newcommand*{\Tx}{T_{\rm X}}
\newcommand*{\SigX}{\Sigma_{\rm X}}

\newcommand*{\trace}{\mathrm{Tr}}
\newcommand*{\gradient}{\nabla}
\newcommand*{\diver}{\nabla\cdot}

\newcommand*{\uv}{\boldsymbol{u}}
\newcommand*{\uphi}{u_{\varphi}}
\newcommand*{\uR}{u_R}
\newcommand*{\uz}{u_z}
\newcommand*{\vv}{\boldsymbol{v}}
\newcommand*{\evphi}{{\bf e}_{\varphi}}
\newcommand*{\vphi}{v_\varphi}
\newcommand*{\vphimax}{v_{\varphi,\mathrm{max}}}
\newcommand*{\vphisqmean}{\overline{\vphi^2}}

\newcommand*{\veldisp}{\boldsymbol{\sigma}}
\newcommand*{\sigmaphi}{\sigma_\varphi}
\newcommand*{\sigmaphimax}{\sigma_{\varphi,\mathrm{max}}}
\newcommand*{\sigmaR}{\sigma_R}
\newcommand*{\sigmaz}{\sigma_z}

\newcommand*{\kint}{k_{\rm int}}
\newcommand*{\kext}{k_{\rm ext}}
\newcommand*{\Rsn}{R_{\rm SN}}
\newcommand*{\therPar}{\gamma_{\rm th}}

\newcommand*{\Msun}{M_{\odot}}
\newcommand*{\Mh}{M_{\rm h}}
\newcommand*{\Mstar}{M_\star}
\newcommand*{\Mhot}{M_{\rm hot}}
\newcommand*{\Mesc}{M_{\rm esc}}
\newcommand*{\Minj}{M_{\rm inj}}
\newcommand*{\alphaSN}{\alpha_{\rm SN}}
\newcommand*{\alphaWinds}{\alpha_\star}

\newcommand*{\nDenH}{n_\mathrm{H}}

\newcommand*{\nden}{n}
\newcommand*{\denstar}{\rho_\star}

\newcommand*{\dtpartial}[1]{\dfrac{\partial#1}{\partial t}}
\newcommand*{\emissivity}{\mathscr{L}}
\newcommand*{\coolFunc}{\Lambda}

\newcommand*{\convective}[1]{\left(#1 \cdot \nabla \right)}
\newcommand*{\norm}[1]{\left\lVert#1 \right\rVert}

\newcommand*{\equanameref}[1]{equation~(#1)\xspace}
\newcommand*{\Equanameref}[1]{Equation~(#1)\xspace}

\newcommand*{\gyr}{\mathrm{Gyr}}

\newcommand*{\mpc}{\mathrm{Mpc}}

\newcommand*{\pc}{\mathrm{pc}}
\newcommand*{\kms}{\mathrm{km~s^{-1}}}
\newcommand*{\enSource}{\mathscr{E}}

\newcommand*{\dd}{\mathrm{d}}

\graphicspath{{./Figures/}}
\pagerange{\pageref{firstpage}--\pageref{lastpage}} \pubyear{2014}

\begin{document}

\title[Stellar dynamics effects on the S0 galaxies X-ray emission]
{The effects of stellar dynamics on the X-ray emission of flat early-type galaxies}
\author[Negri, Ciotti \& Pellegrini]{Andrea Negri\thanks{E-mail:
andrea.negri@unibo.it}, Luca Ciotti and Silvia Pellegrini\\
Department of Physics and Astronomy, University of Bologna, viale Berti Pichat 6/2, I-40127 Bologna, Italy}
\date{Accepted 2013 December 31. Received 2013 November 25; in original form 2013 September 11}

\maketitle
\label{firstpage}

\begin{abstract}
 Past observational and numerical studies indicated that the hot
 gaseous haloes of early-type galaxies may be sensitive to the
 stellar kinematics. With high resolution ZEUS 2D hydro simulations
 we study the hot gas evolution in flat early-type galaxies of fixed
 (stellar plus dark) mass distribution, but with variable amounts of
 azimuthal velocity dispersion and rotational support, including the
 possibility of a (counter)rotating inner disc. The hot gas is fed
 by stellar mass losses, and heated by supernova explosions and
 thermalization of stellar motions. The simulations provide
 $\therPar$, the ratio between the heating due to the relative
 velocity between the stellar streaming and the ISM bulk flow, and
 the heating attainable by complete thermalization of the stellar
 streaming. We find that 1) X-ray emission weighted temperatures and
 luminosities match observed values, and are larger in fully velocity
 dispersion supported systems; X-ray isophotes are boxy where rotation is significant; 2) $\therPar \simeq 0.1-0.2$ for
 isotropic rotators; 3) $\therPar\simeq 1$ for systems with an inner
 (counter)rotating disc. The lower X-ray luminosities of isotropic
 rotators are not explained just by their low $\therPar$, but by a
 complicated flow structure and evolution, consequence of the
 angular momentum stored at large radii.
Rotation is therefore important to explain the lower average X-ray emission
and temperature observed in flat and more rotationally supported galaxies.
\end{abstract}

\section{Introduction}

Early-Type galaxies (ETG) are embedded in a hot ($10^6 -10^7$~K), 
X-ray emitting gaseous halo \citep{fabbiano1989,
 o'sullivan2001}, produced mainly by stellar winds and heated by Type
Ia supernovae (SNIa) explosions and by the thermalization of both
ordered and random stellar motions \citep[e.g., see][]{pellegrini2012}. 
A number of different astrophysical
phenomena determine the X-ray properties of the halo: stellar population
evolution, galaxy structure and internal kinematics, AGN presence, and
environmental effects. A full discussion of the most relevant
observational and theoretical aspects concerning the X-ray haloes can
be found elsewhere \citep[][hereafter
KP12]{mathews.brighenti.2003, Kim.Pellegrini2012}. Among the
questions less understood, there is the role of galaxy shape and
rotation in determining the properties of the hot haloes. In recent
times, \textit{Chandra} observations confirmed the result known since
\textit{Einstein} observations that flattened systems show a lower X-ray
luminosity than rounder systems of similar optical luminosity $\Lb$
\citep{eskridge.et.al.1995, sarzi.et.al.2013}
However, flatter
systems also possess, on average, higher stellar rotation levels, thus
it remains undecided which one between shape and internal kinematics
could be responsible for the observational result \citep[e.g.][]{pellegrini.et.al.1997, sarzi.et.al.2013}. On the theoretical side, despite some
important numerical \citep[][hereafter DC98]{kley.Mathews.1995, brighenti.mathews.1996, dercole.ciotti1998}, and analytical (\citealt{ciottiPellegrini1996}, hereafter CP96; \citealt{posackiproceeding2013, posacki.et.al2013}) works, the situation is still unclear.
Renewed interest in the subject has come
recently after the higher quality \textit{Chandra} measurements of the hot
gas luminosity $\Lx$ and temperature $\Tx$ \citep{boroson.et.al2011}. In an investigation using
\textit{Chandra} and \textit{ROSAT} data for the $\mathrm{ATLAS^{3D}}$ sample, \citet{sarzi.et.al.2013}
found that slow rotators generally have the largest $\Lx$ and $\Tx$;
fast rotators, instead, have generally lower $\Lx$ values, and the
more so the larger their degree of rotational support. The $\Tx$
values of fast rotators keep at $ \lesssim 0.4$ keV, and do not scale with
the central stellar velocity dispersion \citep[see also][]{boroson.et.al2011}.
In this paper we focus on the effects of different amounts
of rotational support on the X-ray properties of the hot haloes, with
an investigation based on hydrodynamical simulations. A more
extensive study, also considering the effects of galaxy shape, will be
done in a subsequent paper.

The effects of rotation are potentially important, and not trivial to
predict from first principles. First, there is their energetic
aspect, since an energy input in the galaxy ISM is associated with the
thermalization of both random and streaming stellar motions. The
thermalization of the random motions provides an energy input per unit
time
\begin{equation}
 \Lsigma\equiv \dfrac{1}{2}\int \mtot\, \trace (\veldisp^2) \dd V, 
\label{eq:Lsigma}
\end{equation}
where $\mtot$ is the total stellar mass injection rate per unit volume
(Sect. 2.2), and $\veldisp$ is the velocity dispersion tensor of the
stellar component\footnote{Hereafter, boldface symbols represent
 vectors and tensors, and $\norm{\cdot}$ is the standard norm.}.
From the energy equation (15) there is an additional
heating contribution due to difference in velocity between the
streaming velocity of the stars ($\vv =\vphi\evphi$) and the ISM
velocity ($\uv$)
\begin{equation}
\Lstr\equiv \dfrac{1}{2}\int \mtot\, \norm{\vv -\uv}^2 \dd V. 
\label{eq:Lstr}
\end{equation}
At variance with $\Lsigma$, this contribution cannot be estimated a
priori, thus we parametrize it by introducing the thermalization
parameter
\begin{equation}
\therPar \equiv \dfrac{\Lstr}{\Lrot},
\end{equation}
where
\begin{equation}
\Lrot\equiv \dfrac{1}{2}\int \mtot\,  \vphi^2 \dd V,
\end{equation}
\citep[see CP96;][]{posackiproceeding2013, posacki.et.al2013}.

From the previous definitions, the total energy transferred to the ISM
per unit time due to stellar motions 
can be written as
\begin{equation}
 \Lkin \equiv \Lsigma + \Lstr = \Lsigma + \therPar \Lrot.
\end{equation}
Of course, $\Lsigma$ decreases when increasing the rotational support
of a galaxy, at fixed galaxy structure. Note that, if $\therPar=1$,
the virial theorem assures that at fixed galaxy structure, $\Lkin$ is
independent of the level of rotational support. In the other extreme
case, if $\therPar = 0$, the gas is injected everywhere with the same
local velocity of the ISM, and then $\Lkin$ decreases for a larger
rotational support. However, ordered rotation acts also in a
competing way, i.e., it tends to unbind the gas; therefore, when
rotation is unthermalized, the ISM is less heated but it is also less
bound. A first important question addressed by the present study is
to obtain estimates of $\therPar$ that can be used in analytical works
\citep[e.g.][]{pellegrini2011, posackiproceeding2013, posacki.et.al2013}.

A second, potentially relevant effect related to ordered rotation,
that can be investigated only with numerical simulations, is the
possibility of large-scale instabilities in the rotating ISM, as those
revealed by the simulations of DC98. The spatial resolution
attainable by DC98, though, was considerably lower than what we reach
in the present study, which is also performed with a different code
and using a different geometry for the numerical grid. These
large-scale instabilities may be at the origin of the well known
observational ``X-ray under-luminosity'' of flat and rotating ETGs mentioned above \citep{eskridge.et.al.1995, sarzi.et.al.2013}.

Finally, we also focus on the role of a counter-rotating stellar disc
that can possibly be present in ETGs (see, e.g., the cases of NGC7097
in \citealt{DeBruyne.et.al2001}, NGC4478 and NGC4458 in \citealt{morelli.et.al2004},
NGC3593 and NGC4550 in \citealt{coccato.et.al2013}, IC719 in \citealt{katkov.et.al2013}, NGC 4473 in \citealt{foster.et.al2013} with a counter-rotating disc
summing up to the 30\% of the total stellar mass, and other cases in
\citealt{kuijken.et.al1996} and in \citealt{erwin.sparke2002}). Here we recognize two
competing effects that could be at work as the ISM flows towards the
centre of a galaxy with a counter-rotating stellar structure. As a
consequence of cooling and conservation of angular momentum, the infalling gas increases its rotational velocity until it reaches the
region where the counter-rotating disc lies, and then interacts with a
counter-rotating mass injection, with the additional heating
(eq. 2). On the other hand, this interaction also causes a reduction
of the specific angular momentum of the local ISM, which will reduce the
local centrifugal support, and will favour central accretion. What of
the two competing effects will dominate can be quantified only with
high resolution numerical simulations. Note that this study
is relevant also for the fuelling of central massive black holes in
rotating galaxies.

In this paper we follow the evolution of the hot ISM in flat ETGs
(modelled as realistic S0 galaxies) with different degrees of
rotational support and dark matter amount, by using the 2D
hydrodynamical code ZEUS-MP2, with the aim of quantifying all the
expected effects described above. In order to reduce the number of
free parameters, the shape of the model galaxy is kept constant for
all the explored models, so that flattening effects are not described
here.

The paper is organized as follows. In Section 2 we describe the
structural and dynamical properties of the galaxy models, and the
input physics. In Section 3 the results are presented describing the
time evolution of several hydrodynamical quantities as a function of
rotational support, estimating the associated $\therPar$, and also
constructing observationally relevant quantities, such as the X-ray
luminosity of the ISM and the emission weighted ISM temperatures.
Finally, in Section 4 the main conclusions are presented.

\section{The Simulations}
\label{the_model}
Numerical simulations of gas flows in ETGs have already been presented
in several previous works (e.g., KP12 for an overview); here we
briefly describe the main ingredients of the present simulations.
Special attention is given only to the implementation of the internal
kinematics of the galaxy models, one of the main ingredients of this
study.

\subsection{The galaxy models}
The galaxy mass profile consists of an axisymmetric stellar model, and a
spherical dark matter halo. The stellar distribution is described by
a \citep{miyamoto.nagai.1975} density--potential pair of total
mass $\Mstar$
\begin{equation}
 \denstar(R,z) = \dfrac{\Mstar b^2}{4 \pi} \dfrac{aR^2 + (a+3\zeta)(a
   + \zeta)^2} {\zeta^3 \left[ R^2 + (a + \zeta)^2  \right]^{5/2}}, 
\label{eq:mndensity}
\end{equation}
\begin{equation}
\Phi_{\star} (R,z) =  - \dfrac{G \Mstar}{\sqrt{R^2 + (a+\zeta)^2}}, 
\label{eq:mnpotential}
\end{equation}
where $a$ and $b$ are scalelenghts, $\zeta \equiv \sqrt{z^2 + b^2}$,
and $(R,\varphi,z)$ are the standard cylindrical coordinates. For
$a=0$, eqs. (6)-(7) reduce to the \citet{plummer.1911} model, while for
$b=0$ to the \citet{kuzmin.1956} disc.

The dark matter (DM) halo is described by a spherical Einasto
density--potential pair of total mass $\Mh$ \citep{einasto,
  Navarro04,Merr,Gao08,Navarro10}:
\begin{equation}
 \rho_{\rm h}(r)=\rho_{\mathrm c} \mathrm{e}^{d_n - x},\quad x\equiv
 d_n \left(\dfrac{r}{\rh}\right)^{1/n},
\end{equation}
\begin{equation}
\Phi_{\rm h}(r)=-\dfrac{G\Mh}{ r}\left[1-\dfrac{\Gamma(3n,x)}{\Gamma(3n)}+\dfrac{
x^n\Gamma(2n,x)}{\Gamma(3n)} \right],
\end{equation}
where $r=\sqrt{R^2+z^2}$ is the spherical radius, $\rh$ is
the half mass radius, $n$ is a free parameter, and $\rho_{\mathrm c} =
\rho_{\mathrm{h}}(\rh) = \Mh d_n^{3n}e^{-d_n}/(4\pi n\Gamma (3n)\rh^3)$. For $d_n$
we use the asymptotic relation
\begin{equation}
d_n\simeq 3n-\frac{1}{3}+\frac{8}{1215~n}
\end{equation}
\citep{RetMon}.

The ordered ($\vv=\vphi\evphi$) and random ($\sigmaphi$ and $\sigma
=\sigmaR=\sigmaz$) velocities of the stellar component are obtained by
solving the Jeans equations under the assumption of a two-integrals
phase-space distribution function, and applying the \citet{satoh1980}
decomposition
\begin{equation}
\vphi = k \sqrt{\vphisqmean - \sigma^2}, \qquad 
\sigmaphi^2 = k^2 \sigma^2 + (1-k^2)\vphisqmean,
\end{equation}
where $k$ is the Satoh parameter, that in the most general case can
vary with position in the meridional plane (CP96). The Jeans
equations are integrated by using a numerical code described in \citet{posackiproceeding2013, posacki.et.al2013}, on a high resolution cylindrical grid, and the results are interpolated via bi-dimensional cubic splines \citep{numericalRecipiesFortran2} on the hydrodynamical grid.

\begin{figure*}
\includegraphics[width=0.8\linewidth, keepaspectratio]{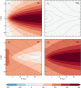}
\caption{Meridional sections of the galaxy rotational field $\vphi$
 (top) and of the stellar azimuthal velocity dispersion $\sigmaphi$
 (bottom) for the \ISi (left) and \VDi (right) models. Note that, by
 construction, the velocity dispersion components $\sigmaR=\sigmaz
 =\sigma$ of the two models coincides with $\sigmaphi$ of \ISi. The
 stellar isodensity contours (solid lines) correspond to a density of
 1 $\Msun~\pc^{-3}$ at the innermost contour, and to a density value
 decreasing by a factor of ten on each subsequent contour going
 outwards.}
\label{fig:fig1}
\end{figure*}
\begin{figure*}
\includegraphics[width=0.8\linewidth,  keepaspectratio]{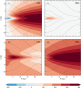}
\caption{Analogue of Fig.~\ref{fig:fig1} for the counter-rotating \CRi model (left),
 and for the velocity dispersion supported model with an inner
 rotating stellar disc, \RDi (right). The rotating inner stellar disc
 is apparent in the top panels.}
\label{fig2}
\end{figure*}

For a given stellar and DM halo mass model, we consider four
different cases of kinematical support for the stellar component: the
isotropic rotator (\IS, $k=1$), the fully velocity dispersion
supported case (\VD, $k=0$), the counter-rotating disc (\CR), and a
velocity dispersion supported system with an inner rotating disc
(\RD). In order to build the CR and RD models, we adopt the
following functional form for the Satoh parameter
\begin{equation}
 k(R,z)= \kext +  \dfrac{\denstar (R,z)}{\denstar (0,0)} (\kint -\kext), 
\label{conter_rot}
\end{equation}
where $\denstar$ is again given by eq. (6), but now $a = 18$ kpc and
$b = 4$ kpc. This choice leads to a very flattened rotating structure
in the central regions of the galaxy, with $k(0,0)=\kint$, while
$k=\kext$ at large radii. In particular, the \CR models are obtained
for $\kint=-1$ and $\kext=1$, while for the \RD models for $\kint=1$
and $\kext=0$. Therefore, the \CR and \RD models at large radii are
similar to the \IS and \VD models, respectively.

\subsection{The input physics}
\label{sec:input_phisics}
The input physics is fully described elsewhere (DC98, KP12), here we
just recall the main points. As usual we account for (1) mass
injection due to stellar mass losses and SNIa ejecta; (2) momentum
sources, due to ordered streaming motions of the stellar component;
(3) energy injection by SNIa explosions and thermalization of random
and streaming stellar motions. The rigorous derivation of the
hydrodynamical equations with isotropic source field\footnote{With
 isotropic source field (in the present case the galaxy stellar
 density distribution, $\denstar$) we mean that the mass, momentum
 and internal energy injected in the ISM by each source element (the
 single stars) are spherically symmetric with respect to the source
 element itself.} is given in DC98. In the present work we solve the
following equations:
\begin{gather}
\dtpartial{\rho} + \diver (\rho\uv) = \mSN + \mWinds
\equiv \mtot 
\label{eq:continuity},\\
\rho\dtpartial{\uv} + \rho
\convective{\uv}\uv = - \gradient p - \rho
\gradient \Phi_{\rm tot} + \mtot\,(\vphi\evphi - \uv) 
\label{eq:euler},\\
\begin{split}
\dtpartial{E} + \diver(E \uv) &= 
- p\diver \uv - \emissivity + \mSN \dfrac{u_s^2}{2} \\
&+ \dfrac{\mtot}{2} \left[ \norm{\vphi\evphi - \uv}^2 +
  \trace (\veldisp^2) \right],
\label{eq:energy}
\end{split}
\end{gather}
where $\rho$, $\uv$, $E$, $p$, $\Phi_{\rm tot}$, and $\emissivity$ are
respectively the ISM mass density, velocity, internal energy density,
pressure, total gravitational potential, and bolometric radiative
losses per unit time and volume. As usual, the gas is assumed to be
an ideal monoatomic fully ionized plasma, so that $p = (\gamma -1)E$
where $\gamma = 5/3$ is the adiabatic index. The chemical composition
is fixed to solar ($\mu\simeq 0.62$), and the gas self-gravity is
neglected. $\mSN=\alphaSN(t) \denstar$, and $\mWinds =\alphaWinds(t)
\denstar$ describe the mass injection rates per unit volume due
respectively to SNIa events and to all kinds of post main-sequence
stellar mass losses; thus:
\begin{equation}
\alphaSN (t) = \dfrac{1.4 \Msun}{\Mstar}\Rsn (t),
\label{eq:SNIa}
\end{equation}
\begin{equation}
\alphaWinds (t) = 3.3 \times 10^{-12} t_{12}^{-1.3} ~(\mathrm{yr^{-1}}),
\label{eq:mass_loss}
\end{equation}
with the SNIa explosion rate $\Rsn (t)$ given by 
\begin{equation}
\Rsn (t) = 0.16\, H_{0,70}^2 \times 10^{-12}\Lb\, t_{12}^{-s} ~ (\mathrm{yr^{-1}}), 
\end{equation}
where $H_{0,70}$ is the Hubble constant in units of 70
$\mathrm{km~s^{-1}~Mpc^{-1}}$, $\Lb$ is the present epoch B-band
galaxy luminosity in blue solar luminosities, $t_{12}$ is the age of
the stellar population in units of $12~\gyr$, and $s$ parametrizes the
past evolution. \Equanameref{\ref{eq:mass_loss}} holds for a Kroupa
Initial Mass Function \citep{pellegrini2012}. The SNIa's heating rate is
obtained as $\Lsn (t)=10^{51}\Rsn (t)$ erg yr$^{-1}$, where $10^{51}$
erg is the kinetic energy of one event. Following recent theoretical
and observational estimates of the SNIa explosion rate
\citep{mannucci.etal2005,greggio2005, greggio2010, sharon.etal2010,
 maoz.etal2011}, we adopted $s=1$. Given that $\Lsn$ is typically
larger than $\Lsigma$ and $\Lrot$ (e.g., Tab.~\ref{tab:simulations}), this choice produces
a long-term time-increase of the specific heating $\Lsn /\mtot$ of the
input mass, due to the different time dependence of the mass and
energy inputs from the evolving stellar populations \citep[e.g.][]{pellegrini2012}.

It may be useful to stress a point not always clear in the discussion
of the energetics of the ISM in ETGs. A generic isotropic source field is
associated with an internal energy source term given by
\begin{equation}
\enSource = \dfrac{\mtot}{2} \left[\norm{\vv -\uv}^2  + \trace (\veldisp ^2) \right] +
\mtot \left( e_\mathrm{inj} + \dfrac{u_s^2}{2} \right),
\end{equation} 
where $\mtot$, $\vv$, $\uv$, $e_\mathrm{inj}$, $u_s$, $\veldisp^2$ are
the mass injection rate per unit volume, the source streaming velocity
field, the velocity of the ambient gas, the internal energy per unit
mass of the injected gas, the modulus of the relative velocity of the
injected material and the source (i.e., the velocity of the stellar
winds and of the SNIa ejecta), and finally the velocity dispersion
tensor of the source field (e.g., DC98). For both the mass sources
considered here the streaming velocity and the velocity dispersion
tensor are the same, so that eqs. (13)-(14) are exact. Some
discussion is instead needed for \equanameref{\ref{eq:energy}}. The
thermalization of random motions is usually neglected in the case of
SNIa's mass input (as well as of the associated internal energy), due
to the high velocity of the ejecta $u_s=\sqrt{2\times 10^{51} {\rm
  erg\, s^{-1}}/1.4 \Msun}\simeq 8.5\times 10^3$ km s$^{-1}$, far
above the typical value of the velocity dispersion in ETGs ($\simeq
150-300 ~\mathrm{km~s^{-1}}$). The opposite applies to stellar winds:
a typical red giant star injects mass in the ISM via winds with a
speed of few $10 ~ \mathrm{km~s^{-1}}$ \citep{parriott.bregman2008},
one order of magnitude lower than the velocity dispersion of a typical
ETG, so that the contribution of the winds internal energy and kinetic
energy is usually ignored. In our work we neglect the $u_s$ term of
stellar winds, but we consider that of SNIa ejecta; this leads to the
present form of eq. (15).
We adopted a thermalization efficiency equal to $0.85$ for the kinetic energy input from SNIa \citep[e.g.][]{thornton.et.al1998, tang.wang2005}.

The radiative cooling is implemented by adopting a modified version of
the cooling law reported in \citet{sazonov2005},
neglecting the Compton heating/cooling and the photo-ionization
heating, allowing only for line and recombination continuum cooling.
We impose a lower limit for the ISM temperature of $T>10^4$~K,
by modifying the cooling function at low temperatures. With these
assumptions, our version of the cooling function, derived from
eq. (A32) of \citet{sazonov2005}, becomes $\emissivity = \nDenH^2
\coolFunc (T) $, where $\nDenH$ is the hydrogen number density and
 \begin{equation}
\coolFunc (T) = \left[ S_1 (T) + 10^{-23}a(T) \right]
\left(1-\dfrac{10^4~\mathrm{K}}{T} \right)^2~(\mathrm{erg~s^{-1}~cm^{-3}}),
\label{cool_function}
\end{equation}
where the $S_1(T)$ and $a(T)$ functions are given in \citet{sazonov2005}.

\subsection{The code}
The simulations are run with the ZEUS-MP~2 code \citep{hayesetal2006},
a widely used Eulerian, operator splitting, fixed mesh, upwind code
which operates in one, two and three dimensions in Cartesian,
spherical and cylindrical coordinates. The code has been modified to
take into account the source terms in eqs. (13)-(15) with a Forward
Time Centered Space (FTCS) differencing scheme.

Due to the ZEUS explicit scheme, the global hydrodynamical time-step
$\Delta t$ takes into account the Courant--Friedrichs--Lewy
stability condition imposing a minimum value $\Delta t_\mathrm{hyd}$
(eq. 60 in \citealt{hayesetal2006}). Our input physics
leads to the introduction of additional characteristic times,
associated with the injection of mass, momentum and energy, and with
radiative cooling:
\begin{equation}
 \Delta t_{\rho}     = \dfrac{\rho}{\mtot},  \quad  
\Delta t_{\rm c} =  \dfrac{E}{\emissivity}, 
\label{eq:timestepmass} 
\end{equation}
\begin{equation}
\Delta t_{\rm h} = \dfrac{2E}{\mSN u_s^2 + \mtot \left[
    \norm{\vphi\evphi - \uv}^2 + \trace (\veldisp^2) 
\right]}, 
\label{eq:timestepenergy}
\end{equation}
so that 
\begin{equation}
\Delta t\equiv \dfrac{C_{\rm cfl}}{\sqrt{\Delta t_{\rm hyd}^{-2}
    +\Delta t_{\rm c}^{-2} + \Delta t_{\rho}^{-2} + \Delta t_{\rm h}^{-2}}},
\label{eq:globalDt}
\end{equation}
where $C_{\rm cfl}$ is the Courant coefficient, and 
the minimum value of $\Delta t$ over the numerical grid is considered.

While almost all the integration of eqs. (13)-(15) is performed by
using an explicit temporal advancement (as prescribed by FTCS), for
the integration of the cooling function we tested two different
numerical algorithms: the fully explicit Bulirsch--Stoer method and the
fully implicit Bader--Deuflhard method
\citep{numericalRecipiesFortran2}. All the results presented in this
work are based on the fully implicit algorithm, since it is far less
computational time-consuming, while giving the same global evolution
of the hot gas flows (as proved with several tests).

The code is used in a pure hydro, 2D axisymmetric configuration with a
non uniform (logarithmic) computational mesh $(R,z)$ of $480\times
960$ gridpoints, having a resolution of $\simeq 90$ pc in the first
$10$ kpc from the centre. Reflecting boundary conditions were set
along the $z$-axis, while on the outer edge of the simulated box the
fluid is free to flow out of the computational grid. We adopted a
cylindrical grid (at variance with DC98, who used a spherical grid) in
order to better resolve the regions near the equatorial plane, where a
cold disc can form. Clearly, such a choice is quite expensive in
terms of computational time, as more gridpoints than in the spherical
case are needed, in order to maintain the shape of the grid reasonably
regular with a logarithmic spacing. We verified by performing several
tests that the code provides an excellent conservation of total mass
and energy, that is given by
\begin{equation}
\int (\enSource -\emissivity +\mtot\Phi) \dd V=
\dfrac{\dd E_{\rm tot}} {\dd t}+\int\left(e +\dfrac{p}{\rho} + \dfrac{\norm{\uv}^2}{2}+\Phi\right)\rho \uv\cdot \boldsymbol{n}\, \dd S,
\end{equation} 
where $e=E/\rho$ is the ISM internal energy per unit mass, $E_{\rm
  tot}=\int (e+||\uv ||^2/2+\Phi)\rho \dd V$, and the two integrals are
extended over the whole numerical grid and its boundary, respectively.
During the whole evolution, an amount of gas mass is lost out of the
grid that is comparable to, or within a factor of few larger than, the
present-epoch hot gas mass (Tab.~\ref{tab:simulations}).

The hydrodynamical fields are saved every $100$ Myr, while
grid-integrated quantities, such as the cumulative injected mass by
the evolving stellar population ($\Minj$, stellar winds plus SNIa
ejecta), the cumulative mass escaped from the galaxy ($\Mesc$), the
hot gas mass ($\Mhot$, having $T \geq 10^6$ K), $\Lstr$, $\Lrot$,
$\Lsigma$ (eqs. 1-4), the X-ray emission in the 0.3--8
keV \textit{Chandra} band $\Lx$, and the X-ray emission weighted
temperature ($\Tx$), are sampled with a time resolution of 1
Myr. $\Lx$ and $\Tx$ are calculated using the thermal emissivity
$\emisx$ over 0.3--8 keV emission of a hot,
collisionally ionized plasma, using the spectral fitting package
\textsc{XSPEC}\footnote{http://heasarc.nasa.gov/xanadu/xspec/.} \citep[spectral
model \textsc{apec},][]{smithetal2001}. Thus:
\begin{equation}
\Lx = \int \emisx \dd V, \quad \Tx = \dfrac{ \int T \emisx \dd V}{\Lx},
\label{eq:Tx} 
\end{equation}
where the integration extends over the whole computational grid.
Finally, the X-ray surface brightness maps $\SigX$ were also
constructed for an edge-on projection, where the rotational and
flattening effects are maximal.

\section{Results}\label{results}
We present here the main results of our investigation, focussing
on a representative selection of models. The detailed features of each
simulated flow of course depend on the specific galaxy model and input physics.
While the parameter space is too large for a complete
exploration, fortunately, the global behaviour of the gas is quite robust against minor changes of the input parameters; thus, a
reasonable amount of computational time is sufficient to capture the
different behaviour of the flows resulting from major variations in the
structural parameters of the parent galaxy.
\begin{table}
\begin{center}
\caption{Main outputs at the end of the simulations (13 Gyr).}\label{tab:simulations}
\begin{tabular}{cccccccc}
\toprule
Name       &  $\Mhot$ & $\Mesc$ & $\Lsigma$ & $\Lrot$ & $\Lx$ & $\Tx$ \\
\midrule                                             
\ISi     &    1.17  & 2.21 & 1.85 & 1.46  & 0.58   & 0.40\\   
\VDi     &    1.35 & 2.40 & 3.31 &  0.00   & 3.38  & 0.50 \\
\CRi    &    1.38  & 2.53 & 2.58 & 0.73 & 1.03   & 0.40\\
\RDi    &    1.36  & 2.50 & 3.13 & 0.18 & 1.99  & 0.55\\
\midrule                                   
\ISl     &    1.27  & 4.28  & 1.54 & 1.06 & 0.41 & 0.32\\   
\VDl     &    1.17 & 4.50  & 2.60 &  0.00  & 3.08 & 0.39\\
\CRl    &    1.01  & 4.90  & 2.10 & 0.50 & 0.14 & 0.37\\
\RDl    &    1.05  & 4.71 & 2.47 & 0.14 &  1.21 & 0.42\\
\midrule   
\ISh     &    1.16  & 1.28 & 2.46 & 2.28 & 0.24   & 0.55\\   
\VDh     &    1.68 & 1.37 & 4.74 &  0.00   & 6.40 & 0.70\\
\CRh    &    1.52  & 1.42 & 3.55 & 1.20 & 1.60  &  0.55 \\                                         
\RDh    &    1.54  & 1.43  & 4.48 & 0.26 & 2.63 & 0.77 \\
\bottomrule
\end{tabular}
\end{center}
\textit{Notes.} The columns give the model name, 
 the hot ISM mass within the computational grid, the escaped mass from the grid boundary, $\Lsigma$ (eq. 1) and
 $\Lrot$ (eq. 4), the ISM 0.3--8 keV luminosity and the emission
 weighted temperature (eq. 25). Masses are in units of $10^9\Msun$,
 luminosities in $10^{40}$ erg s$^{-1}$, and $\Tx$ in keV. For
 reference, at 13 Gyr the total mass injected in the galaxy from the
 beginning by the evolving stellar population (stellar winds plus SNIa
 ejecta) is $\Minj=2.23\times 10^{10} \Msun$, and the SNIa's heating
 rate is $\Lsn =1.5\times 10^{41}$ erg s$^{-1}$.
\end{table}

In all models the stellar distribution is kept fixed (the effects of a
variation in the galaxy shape is studied in a subsequent work, Negri
et al. in preparation). For reference we adopt a galaxy model tailored
to reproduce the main structural properties of the Sombrero galaxy
(M~104, of morphological type Sa), taken as a representative case of a flat and rotating
galaxy; at this stage, though, we are not concerned with
reproducing in detail the properties of the X-ray halo of Sombrero
(but see Sect. 3.4). The stellar mass of Sombrero is
$\Mstar\simeq 2.3 \times 10^{11}\Msun$ \citep{tempel.tenjes.2006}.
When adopting Sombrero's apparent blue magnitude of 8.98
\citep{1991rc3}, and a distance of $9.8~\mpc$
\citep{jardel.el.al.2011}, the resulting $B$-band luminosity is
$\Lb\simeq 3.8\times 10^{10} L_{\rm B,\odot}$. In order to reproduce
the major photometric and kinematical features of M~104 as given by
\citet{jardel.el.al.2011}, under the assumption that the galaxy is an
isotropic rotator, we fixed $a=b=1.6$ kpc in eqs. (6)-(7), and $n=4$
and $\rh = 52.8$ kpc in eqs. (8)-(9). The resulting DM halo is
characterized by a total mass of $\Mh = 2.5 \times 10^{12}\Msun$ and
$\rho_c = 4.96\times 10^{-26}$ g cm$^{-3}$. We call this model \ISi,
where the subscript ``$\rm {i}$'' stands for ``intermediate halo'',
for reasons that will be clear in the following. The kinematical
fields of model \ISi are given in Fig.~\ref{fig:fig1} (left panels), superimposed
on the isodensity contours of the stellar distribution. For reference,
in the equatorial plane the maximum stellar streaming velocity is
$\vphimax\simeq 396$ km s$^{-1}$ at $R\simeq 3$ kpc, and $\sigma =
245$ km s$^{-1}$ at the centre.

Starting from \ISi, we built three more models characterized by a
different internal kinematics, but with the same stellar and DM halo
distributions. In model \VDi all the galaxy flattening is supported by
azimuthal velocity dispersion [$k=0 $ in eq. (11); Fig.~\ref{fig:fig1}, right
panels]; in the equatorial plane $\sigmaphimax\simeq 401$ km s$^{-1}$
at $R\simeq 12$ kpc. In the counter-rotating model \CRi ($\kint=-1$
and $\kext=1$ in eq. 12), the equatorial negative and positive
rotational velocity peaks are $-155$ km s$^{-1}$ and $377$ km
s$^{-1}$, reached at $R\simeq 2$ kpc and $R\simeq 24$ kpc,
respectively, while the circle of zero rotational velocity is at
$R\simeq 4.9$ kpc. In practice, \CRi is similar to \ISi in the
external regions, but has a thin counter-rotating stellar disc in the
inner region (Fig.~\ref{fig2}, left panels). For this model $\sigmaphimax\simeq
370$ km s$^{-1}$ at $R\simeq 5.5$ kpc. Finally, in the \RDi model an
inner stellar rotating structure is present ($\kint=1$ and $\kext=0$
in eq. 12) with $\vphimax\simeq 200$ km s$^{-1}$ at $R\simeq 3$ kpc,
while at large radii the galaxy flattening is supported by the
velocity dispersion, similarly to what happens for \VDi (Fig.~\ref{fig2}, right
panels). Note that in all these models, by construction, the velocity
dispersion fields $\sigma=\sigmaR=\sigmaz$ are the same as in model
\ISi (being the Jeans equation along the $z$ axis unaffected by the
amount of ordered azimuthal motions), and coincide with the field
$\sigmaphi$ of \ISi.

In addition to these four models, hereafter referred to as having an
intermediate DM halo mass, we built two more groups of models, where
the DM mass is doubled with respect to the intermediate halo ones
($\Mh = 5 \times 10^{12} \Msun$; hereafter ``heavy halo'' models \ISh,
\VDh, \CRh and \RDh), and where the dark mass is halved ($\Mh = 1.25
\times 10^{12}\Msun$; hereafter ``light halo'' models \ISl, \VDl, \CRl
and \RDl). In the heavy and light halo models again the stellar
distribution is kept fixed, as also $n$ and $\rh$; in addition the four
choices for the $k(R,z)$ field corresponding to the IS, VD, CR and RD
internal kinematic pattern are maintained (see Tab.~\ref{tab:par_models}).
Summarizing, we followed the ISM evolution for a set of 12 models; a few more models
with ``ad hoc'' modifications in the input physics have been also
run, in order to test specific issues, as discussed in the
following Sections.

As usual in similar studies, the galaxy structure and dynamics are kept fixed during the simulations,
and the initial conditions assume the galaxy is devoid of gas, as 
expected after the period of intense star formation, giving birth to the
galaxy, is ended by the strong feedback from type II supernovae. In this way, 
simulations do not start with an equilibrium ISM configuration, but instead
the hot ISM distribution builds-up from stellar mass losses with time increasing.
All the simulations start at an initial galaxy age of 2
Gyr, and the evolution of the gas flow is followed for 11 Gyr. Merging
and gas accretion from outside are not considered; star formation and
black hole feedback are also ignored.

\subsection{Hydrodynamics}
All models, independently of their internal dynamics, evolve through
two well defined hydrodynamical phases. Initially, all the ISM
properties are characterized by an almost perfect symmetry with
respect to the galaxy equatorial plane ($z=0$). As time increases, the
specific heating of the stellar mass losses increases (Sect. 2.2), and
the velocity field becomes increasingly structured, in a way that is
related to the particular internal kinematical support of the stellar
component, as described below. A time arrives when the reflection
symmetry is lost, and from this moment on it is never restored. In the
following we present the main characterizing features of the gas flows
in \VD, \IS, \CR and \RD models. A summary of the relevant integrated
quantities at the end of the simulations is given in Tab.~\ref{tab:simulations}.
\begin{table}
\begin{center}
%
\caption{Stellar kinematics of the models.}\label{tab:par_models}
\begin{tabular}{ccccccccccc}
\toprule
Name       &  $k$& $k_\mathrm{int}$& $k_\mathrm{ext}$  &  $M_\mathrm{rot}$ & $M_\mathrm{crot}$  & $\vphimax$ & $\sigmaphimax$ & $J_z$\\
\midrule                                             
\ISi     &    1  & -- & -- & 1.00& -- & 396 & 245   &31.2\\
\VDi     &    0  & -- & -- & 0.00& -- & 0.00& 401   &0.00   \\
\CRi     &    -- & -1 &  1 & 0.72&0.28& 378 & 370   &24.4\\
\RDi     &    -- & 1  & 0  & 1.00&0.00& 200 & 400   &3.44\\
\midrule                                             
\ISl     &    1  & -- & -- & 1.00& -- & 332 & 223   &25.5\\   
\VDl     &    0  & -- & -- & 0.00& -- &0.00 & 340   &0.00   \\
\CRl     &    -- & -1 &  1 & 0.72&0.28& 303 & 327   &19.5\\
\RDl     &    -- & 1  & 0  & 1.00&0.00& 178 & 338   &2.99\\
\midrule                                             
\ISh     &    1  & -- & -- & 1.00& -- & 513 & 284   &40.3\\   
\VDh     &    0  & -- & -- & 0.00& -- & 0.00& 516   &0.00   \\
\CRh     &    -- & -1 &  1 & 0.72&0.28& 499 & 445   &31.9\\
\RDh     &    -- & 1  & 0  & 1.00&0.00& 239 &  515  &4.18\\
\bottomrule
\end{tabular}
\end{center}
\textit{Notes.} For each model the columns give: the $k$ Satoh parameter of IS and VD models, the parameters $\kint$ and 
$\kext$ in eq. (12) for CR and RD models, the rotating ($M_\mathrm{rot}$) and counter-rotating ($M_\mathrm{crot}$) stellar mass normalized to $\Mstar$, the maximum values of the stellar streaming velocity and of the azimuthal velocity dispersion
in $\kms$, and the total angular momentum of the stars in $10^{73}$ g cm$^2$ s$^{-1}$.
\end{table}

\subsubsection{VD models}
\begin{figure*}
\includegraphics[width=1.0\linewidth, keepaspectratio]{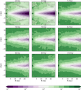}
\caption{Hydrodynamical evolution of the \VDi model, for a selection
 of representative times. Arrows show the meridional velocity field
 $(\uR, \uz)$ of the ISM; their lenght is proportional to the modulus of the gas velocity,
and is normalized to the same constant value in each one of the nine panels, 
and in all panels of the subsequent Figs.~\ref{fig4}-\ref{fig10}; thus,
the evolution of the velocity field as a function of time can be followed for
a single model, and compared to that of the other models in Figs.~\ref{fig4}-\ref{fig10}. 
For reference, the longest arrow in the top left panel corresponds to
171 km s$^{-1}$. 
Colours map the ratio of heating over cooling times, $\Delta t_{\rm h}/\Delta t_{\rm c}$, 
as defined in eqs. (21) and (22); green and violet colours indicate
 cooling and heating regions, respectively.}\label{fig3}
\end{figure*}

\begin{figure*}
\includegraphics[width=1.0\linewidth, keepaspectratio]{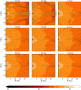}
\caption{ISM temperature evolution for model \VDi, at the same times as in Fig.~\ref{fig3}.
The arrows indicate the velocity field in the meridional plane, and are normalized 
as in Fig.~\ref{fig3}.
The colour-bar indicates the temperature values in K.} 
\label{fig4}
\end{figure*}

We present here the time evolution of the ISM of the non-rotating, fully
velocity dispersion supported models \VDi, \VDl and \VDh. Snapshots of
various flow properties in the meridional plane $(R,z)$ for the \VDi
model, for a selection of 9 representative times, are shown in Figs.~\ref{fig3}
and \ref{fig4}. In particular, in Fig.~\ref{fig3} the colours map the ratio $\Delta
t_{\rm h}/\Delta t_{\rm c}$ (eqs. 21-22), with green and violet
corresponding to cooling and heating regions, respectively, while in
Fig.~\ref{fig4} we show the ISM temperature field for the same times. In both
figures, the arrows represent the ISM meridional velocity field $(\uR,
\uz)$.

The major feature characterizing the flow of \VDi is present from the
beginning of the evolution: this is the degassing along the galaxy
equatorial plane, due to the concentrated heating there, and accretion
on the galaxy centre along the $z$-axis. Above the plane, on a scale
of $\simeq 10$ kpc, the flow is characterized by large-scale
regular vortices (a meridional circulation). Due to the lack of
centrifugal support, cold gas accumulates at the centre from the
beginning. Loss of reflection symmetry of the flow occurs at $t\simeq
4.5$ Gyrs. After this time little evolution takes place, and overall
the gas velocity field slowly decreases everywhere. The flow remains
decoupled kinematically: the axial inflow-equatorial outflow mode
persists in an essentially time-independent way for the entire run,
with heated and outflowing ISM in the disc, and almost stationary gas
above and below the galactic disc.

The ISM temperature, after an initial phase in which the gas is hotter
in the outflowing disc, quickly establishes on a spherically
symmetric structure (Fig.~\ref{fig4}). From the beginning, at the centre
(within $\simeq 500$ pc), the gas cools and forms a dense cold
core. Outside this region, the temperature is steeply increasing,
forming (within $\simeq 1$ Gyr) a spherical, hot region ($T\simeq
9\times 10^6$ K at the peak), of radius $r\simeq 5$ kpc; at larger
radii, the temperature is slowly decreasing outward, keeping a
spherical distribution. Outside the central cool core, the temperature
is everywhere slowly increasing with time, due to the secular increase of the specific ISM heating due to the adopted SNIa
time evolution (eqs. 16-18).

The major features described above for the \VDi model are
qualitatively independent of the DM halo mass, although important
trends with $\Mh$ are clearly detected. For example, the time of loss
of reflection symmetry in the flow properties increases from $\simeq
4$ Gyr (\VDl) to $\simeq 4.5$ Gyr (\VDi) to $\simeq 5.2$ Gyr (\VDh).
In addition, a more massive DM halo tends to
``stabilize'' the ISM velocity field, in the sense that in the
light-halo model \VDl the meridional vortices are more pronounced, and
the temperature maps (while still showing a spherically symmetric
structure on average) are more structured and less regular. In
particular, the average values of the ISM velocity are higher (at any
given time) for lighter DM haloes, and the equatorial violet region in
figures analogous to Fig.~\ref{fig3} (not shown) is less symmetric. Finally, at
any time the average ISM temperature is larger for increasing $\Mh$.

\subsubsection{IS models}
The evolution of the flow in the family of isotropic rotators is
more complicated than in \VD models, as already found in DC98, albeit
for different galaxy models and different input physics.
Figures 5 and 6 show the flow properties of the \ISi model, at the same epochs of
Figs.~\ref{fig3} and \ref{fig4} for the \VDi model. The only similarities with \VDi are
the loss of reflection symmetry, that happens at now at $\simeq 3.3$
Gyr, and a systematic decline in the flow velocity for increasing
time. Noticeable differences are instead apparent. First of all, in
\ISi there is the formation, since the beginning, of a cold and thin
gaseous rotating disc in the inner equatorial galaxy region (with size
$\simeq 5$ kpc), due to angular momentum conservation. This cold disc
is quite stable, even though cooling instabilities from time to time
lead to the formation of cold blobs detached from it. In general, the
cooling (green) regions are significantly more rich in substructures
than in \VDi.
In particular, a second major difference with respect to
\VDi is given by the presence of a cooling $V$-shaped region containing
the equatorial plane whose vertex matches the outer edge of the rotating
cold disc.
In this region the gas is colder than in
the rest of the galaxy (except for the cold rotating disc;
see for example the snapshots at 2.7, 8.6 and 13 Gyr in Fig.~\ref{fig6}).
This V-region becomes cyclically more or less prominent
during the evolution; when it is more prominent, the gas in it is
almost at rest in the meridional plane (i.e., it is fully supported by
its rotational velocity $\uphi$). Inside the $V$-shaped region, the
gas is outflowing along the equatorial disc, while outside the ISM
velocity field is organized in large meridional vortices. Note how
this $V$-shaped region nicely maps the region of similar shape in
Fig.~\ref{fig:fig1} (bottom left panel), where it is clear how the heating
contribution from the thermalization of the stellar azimuthal velocity
dispersion is missing with respect to the \VDi model.

\begin{figure*}
\includegraphics[width=1.0\linewidth, keepaspectratio]{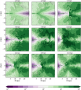}
\caption{Hydrodynamical evolution of the \ISi model at the same
 representative times of model \VDi.} 
\label{fig5}
\end{figure*}

\begin{figure*}
\includegraphics[width=1.0\linewidth, keepaspectratio]{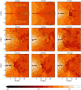}
\caption{ISM temperature evolution of the \ISi model, at the same
 times as in Fig.~\ref{fig5}.} 
\label{fig6}
\end{figure*}

A third major difference between \ISi and \VDi is represented by the
long-term time evolution of the heated (violet) regions in Fig.~\ref{fig5}. In
fact, in \ISi it is apparent the fading of the heated equatorial disc
region, accompanied by the appearance of a central heated region. In
the \VDi model, instead, cooling always prevails over
heating in the centre. This difference is due to the lower gas density in the
central regions of \ISi with respect to \VDi, which is produced by the
angular momentum barrier of the \ISi model, that prevents the gas from
falling directly into the central galactic region. Moreover, in \ISi,
the infalling gas accumulates on the cold disc, and this further
decreases the hot gas density in the central galactic region with respect to \VDi. Thus, in the central galactic region, the
cooling time keeps shorter in \VDi than in \ISi, during their secular
evolution. This difference in the central gas density between
rotating and non-rotating models, with the consequent secular heating
of the central gas in \ISi\footnote{Recall that in all these models
 the specific heating is increasing with time.}, and the constant
cooling of that in \VDi, is at the base of a fourth major difference
in the respective gas evolutions: the evolution is quite smooth in \VD
models, while it shows a cyclic behaviour in \IS ones, as apparent from
the panels relative to $7.3-8.6$ Gyr in Fig.~\ref{fig5}, which describe a full
cycle (a new cycle starts at $t=8.9$ Gyr; in Sect. 3.3 we describe the
evolution of other gas properties during a cycle). At the beginning of
a cycle ($t=7.3$ Gyr in Fig.~\ref{fig5}), the ISM in a central and almost
spherically symmetric region becomes hotter and hotter, which produces
a pressure increase in this region.
This pressure increase causes an
outflow from the centre, along the disc; as a consequence, the gas
residing at $R\simeq 10$ kpc is compressed, increasing its density and
lowering its temperature (Fig.~\ref{fig6}).
The regular shape of the $V$-region
is disrupted: some of the centrally outflowing gas breaks into its
vertex, and succeeds in reaching out along the disc ($t=7.6$ and $7.8$ Gyr); some other gas
circulates above and below the disc, in complex meridional vortices,
compressing the gas there. 
In \ISi the heating is not strong
enough to establish a full and permanent degassing; the
compression increases the cooling, until a maximum in the extension of
the cooling (and of the low temperature) regions is reached, after
which the flow reverts to the ``original'' state (e.g., shown by
$t=8.9$ Gyr in Figs.~\ref{fig5} and \ref{fig6}). These periodic changes in the ISM
structure are mirrored in the evolution of $\Lx$, as discussed in
Sect. 3.3.

The ISM temperature distribution of \ISi is shown in Fig.~\ref{fig6}. The major
characteristic features of the hydrodynamical evolution are
apparent. The cold thin rotating disc is visible, together with the
embedding spherical region of hotter gas: it is interesting to note
how the disc size corresponds to the radial extent of the hot region.
The cold disc is dense (number density $\nden \gtrsim
10~\mathrm{cm^{-3}}$) and azimuthally supported by ordered rotation,
with peak values of $\uphi = 420~\mathrm{km \, s^{-1}}$. The presence
of the hot, spherical region, with radius matching that of the disc,
is due to the efficient way in which gas cools and joins the disc;
this depletes the central galactic region of gas, thus the heating of
the remaining gas is more efficient. In fact, within $\simeq 3$ kpc
from the centre, the average gas density is $\approx 10^{-2}$
cm$^{-3}$ (or less) in the \ISi model, and $\approx 10^{-1}$ cm$^{-3}$
(or more) in the \VDi model, at least an order of magnitude larger.
Also, within the same radius, excluding the cold core (for the
\VDi) and the cold disc (for the \ISi), the average temperature is
$\simeq 10^7$ K for the \ISi, and lower ($5\times 10^6$ K) for the
\VDi (Figs.~\ref{fig4} and \ref{fig6}).
Outside the central hot sphere, though, the temperature of
\ISi is everywhere lower than that of the \VDi model, and the ISM is centrifugally supported by ordered
azimuthal velocity.

Maps of the Mach number show 
that the ISM velocity field is in general subsonic over the whole galaxy body.
As a consequence, the X-ray emission is not associated with shocks; instead, inhomogeneities in the ISM usually 
cools more effectively (as those in the V-shaped region), and contribute to the total X-ray emission\footnote{
Empirical 
evidence for the absence of shock-related X-ray emission is given in Sect. 3.3, where VD models are shown to
be the most X-ray luminous, while showing the less structured velocity pattern. Cooling inhomogeneities instead affect 
the secular trend of the X-ray emission in IS models, never being able, though, to make rotating models more 
X-ray luminous than non-rotating ones (Sect. 3.3). }
(see also Sect. 3.3 where $\Lx$ and $\Tx$ of all models are discussed).

As for the \VD models, a decrease in DM halo mass causes the ISM
temperature overall to decrease, and the density and velocity fields
become more and more rich in substructures. Due to the lower
importance of angular momentum (a consequence of the reduction of the
ordered stellar streaming velocity), the size of the cold disc
decreases from $\simeq 10$ kpc (\ISh), to $\simeq 5$ kpc (\ISi), to
$\simeq 4$ kpc (\ISl). Remarkably, the size of the hot spherical
region is always the same as the size of the cold disc. The ISM
rotational velocities in the $V$-shaped region also decrease for
decreasing DM halo mass; instead, at late times, \ISl presents a polar
outflow, with velocities of the order of $\uz \simeq 250$ km s$^{-1}$
at $\simeq 10$ kpc above the equatorial plane. These polar outflows
are significantly reinforced in test models in all similar to \ISi,
but with doubled SNIa rate. A change in DM halo mass has also some
complex consequences, coming from the
interplay between heating and binding energies at the galactic centre.
The main characteristics of the global evolution of \ISi and \ISl are
very similar, while the cyclic behaviour is far less prominent in \ISh,
and becomes almost absent after a few Gyr of evolution (see also
Sect. 3.3). In fact, being the central potential well deeper in \ISh
than in the other models, the outflow velocity of the disc gas is
lower, the effect of compression of the surrounding gas is also lower,
and major cooling episodes, with the associated substructure in the 
flow density and velocity patterns, are absent.

\subsubsection{CR models}
\begin{figure*}
\includegraphics[width=1.0\linewidth, keepaspectratio]{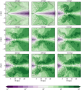}
\caption{Hydrodynamical evolution of the counter-rotating \CRi model, at the same representative times
of \VDi .} 
\label{fig7}
\end{figure*}

We now focus on the first of the two special families of rotating
galaxy models, namely the counter-rotating ones (\CR). As described in
Sect. 2, counter rotation is introduced in the \IS family adopting a
convenient functional form for the coordinate-dependent Satoh
parameter; in particular, we constructed the counter rotation so that
$\vphi =0$ at $R\simeq 5$ kpc, i.e., at the edge of the region where
\ISi develops the cold rotating disc.
As anticipated in Sect.~1, this choice maximizes the possible effects of
counter rotation, both from the energetic and the angular momentum
points of view. As can be seen from Fig.~\ref{fig7}, the global behaviour of
the \CRi model is somewhat intermediate between those of the \VDi and
\ISi models: in fact, although the $V$-shaped cooling region is still
present, it is quite reduced with respect to that in \ISi
(even in regions where counter rotation does not have a direct effect),
and a stronger galactic disc outflow along the equatorial plane takes
place. Correspondingly, the cold gaseous central disc is smaller (with
maximum size of $\simeq 3$ kpc), a consequence of the combined effects
of a stronger local heating and the decrease of the local angular
momentum of the ISM due to the mass injection of the counter-rotating
stellar structure. Overall, however, the hydrodynamical evolution is
similar to that of \ISi, showing that the reservoir of angular
momentum at large radii (where \CRi and \ISi are identical by
construction) is the leading factor in determining the flow behaviour.
In
particular, the \CRi velocity field is more rich in substructures than
in \VDi.

\begin{figure*}
\includegraphics[width=1.0\linewidth, keepaspectratio]{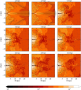}
\caption{Temperature evolution of the counter-rotating \CRi model, at the same representative times
of Fig.~\ref{fig7}.} 
\label{fig8}
\end{figure*}

The temperature evolution of \CRi is presented in Fig.~\ref{fig8}. Again, as in
\ISi, the size of the cold disc strictly matches the size of the
spherical region of hotter gas embedding the cold disc itself. In the
azimuthal velocity field of the ISM, counter rotation is present at
early times, when stellar mass losses are more important; this
produces a region of hotter gas that is not present in the \ISi model,
and which is apparent at the radius of maximum stellar counter rotation
in the first two temperature maps ($t=2.5$ and 2.7 Gyr in
Fig.~\ref{fig8}, cfr. with corresponding panels in Fig.~\ref{fig6}), as a
lighter-coloured area along the equatorial plane, starting from $\simeq
2$ kpc. As time increases, however, the relative importance of the
injected counter-rotating gas decreases, and the rotational velocity of
the ISM becomes dominated by the angular momentum of the ISM inflowing
from the outer regions, and gas counter rotation is no
longer present.

A variation of the DM halo mass leads to the same systematic trends of
the other families: the global ISM temperature decreases for
decreasing $\Mh$, the density and velocity fields become more
structured, the linear size of the cold disc decreases (from 12 to 5
to 3 kpc in radius), and the gas rotational velocity in the $V$-shaped
region decreases. Remarkably, a sustained polar outflow with
$\uz\simeq 300$ km s$^{-1}$ at a height of $\simeq 10$ kpc above the
equatorial plane develops in \CRl by the present epoch, similarly to
what happens for \ISl.

\subsubsection{RD models}
\begin{figure*}
\includegraphics[width=1.0\linewidth, keepaspectratio]{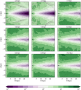}
\caption{Hydrodynamical evolution of the velocity dispersion supported
 model with rotating inner disc, \RDi, at the same representative
 times of\VDi .} 
\label{fig9}
\end{figure*}

We conclude with the family of the \RD models, that are similar to the
\VD ones except for the presence of a rotating stellar disc in their
inner equatorial region. Note that the rotating disc is the only
source of angular momentum in this family. The hydrodynamical
evolution of \RDi is summarized in Fig.~\ref{fig9}, where the global
similarities with \VDi are apparent. In particular, at variance with
\ISi and \CRi, the $V$-shaped region is now missing, while a small
($\simeq 2$ kpc radius) cold disc is present, originated ``in situ''
by the stellar mass losses in the inner rotating stellar disc. The
equatorial outflow is still present as in \VDi, and the ISM velocities
decrease as time increases. Also, in analogy with \VDi and at variance
with \ISi and \CRi, the central spherical heating region does not
appear at late times. This shows again how the global evolution of a
model is strictly linked to the amount of angular momentum stored at
large radii, more than to the specific rotation in the central
galactic regions.

\begin{figure*}
\includegraphics[width=1.0\linewidth, keepaspectratio]{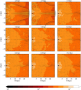}
\caption{Temperature evolution of the velocity dispersion supported
 model with an inner rotating stellar disc, \RDi, at the same times of Fig.~\ref{fig9}.} 
\label{fig10}
\end{figure*}

The temperature evolution of \RDi is shown in Fig.~\ref{fig10}, where again the
similarities with \VDi are apparent. In particular, the temperature
field is much less structured than in the rotating \ISi and \CRi
models. As expected, almost no ISM azimuthal rotation is present in
\RDi, with the exception of some degree of rotation confined in the
inner regions, where the small cold disc resides. The disc is embedded
in a spherically symmetric region of hotter gas, similar to that
present on a larger scale in the \ISi model, and due to the same
cause.

The variation of the DM halo mass leads to the same overall
changes as in the other families: at any given time, for decreasing
$\Mh$, the ISM temperature is lower and the hydrodynamical fields are
systematically less regular, with higher average outflow velocities in
the equatorial plane of the galaxy, while the extension and the
rotational velocities of the small inner cold region decreases.

\subsection{The thermalization parameter}
As discussed in Sect.~1, one of the main goals of this work
is to measure the thermalization parameter $\therPar$ (eq. 3), i.e.,
to estimate how much of the kinetic energy associated with ordered
rotation of the stellar component is converted into internal energy of
the ISM (eq. 2). In fact, in addition to the obvious physical
relevance of the question, reliable estimates of the value of
$\therPar$ as a function of the galaxy rotational status are useful in
theoretical works (e.g., involving estimates of $\Lx$ and $\Tx$ based
on energetic considerations, without simulations; CP96; \citealt{pellegrini2011, posackiproceeding2013, posacki.et.al2013}).
%
\begin{figure*}
\includegraphics[width=\linewidth, keepaspectratio]{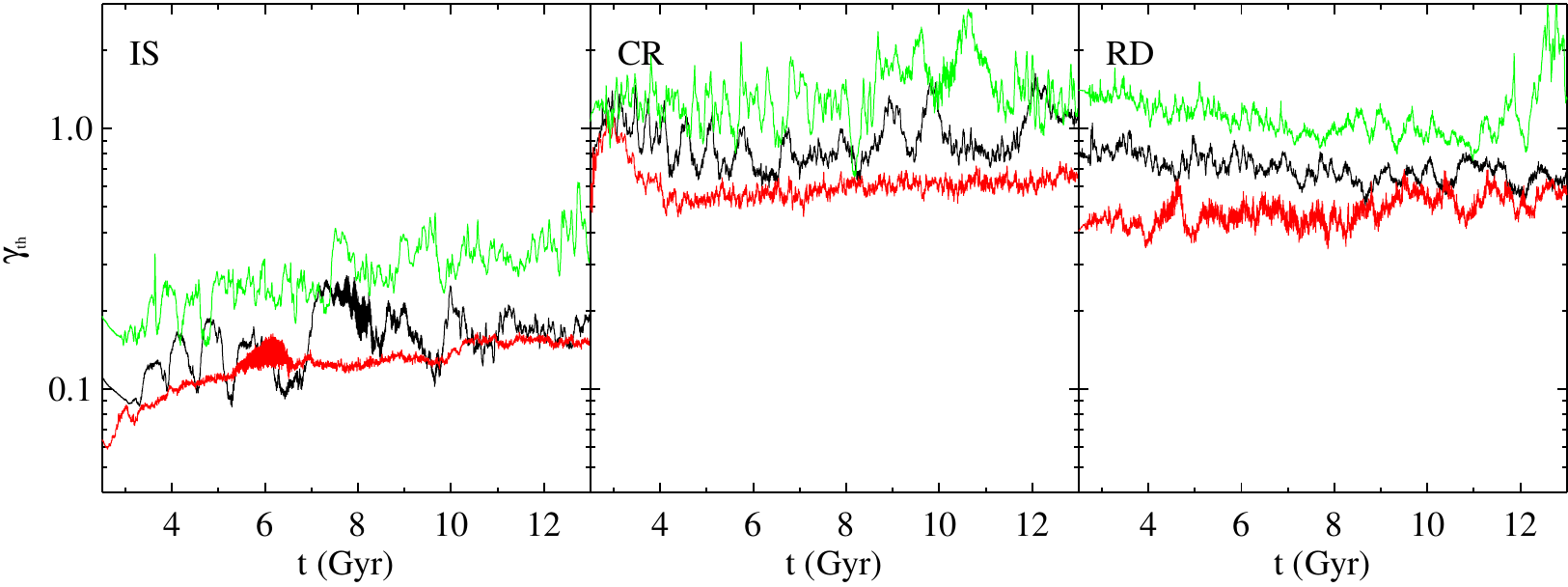}
\caption{Time evolution of the thermalization parameter $\therPar$ for
 the three families \IS, \CR, and \RD: heavy, intermediate, and light
 DM haloes are shown with red, black, and green lines, respectively.} 
\label{fig11}
\end{figure*}

The summary of the results for the three families \IS, \CR, and \RD is
given in Fig.~\ref{fig11}, where red, black, and green lines give the
$\therPar$ values for high, intermediate, and light DM haloes,
respectively. The \VD family is not considered, being the associated
$\therPar$ undefined (formally infinite, being $\Lrot =0$ and $\Lstr
=0.5\int\mtot ||\uv ||^2 \dd V$).
Note that in principle $\therPar$ can
be even larger than unity for galaxies with low rotation and thus low $\Lrot$ (as the \RD models), or in cases of substantial
counter rotation with high $\Lstr$ (as for \CR models). A preliminary
study \citep{negri.et.al.2013proceeding} indicated that $\therPar$ can be quite small in
isotropic rotators.

From Fig.~\ref{fig11} a few common trends are apparent, that can be easily
explained when considering the hydrodynamical evolution of the
models. The first is that for all IS, CR and RD models the value of
the thermalization parameter decreases for increasing DM halo mass. As
$\therPar=\Lstr/\Lrot$, this can be explained by the combination of
two effects: the increase of $\Lrot$ with $\Mh$, coupled with the
decrease of the ISM velocity in the meridional plane and the increase
of the azimuthal component of the ISM velocity. In fact, while the
effect of $\Mh$ on $\Lrot$ is obvious, the effect on $\Lstr$ becomes
clear when recasting eq. (2) as
\begin{equation}
\Lstr = \dfrac{1}{2}\int\mtot\, (\uR^2+\uz^2) \dd V+ 
\dfrac{1}{2}\int\mtot\, (\vphi-\uphi)^2 \dd V, 
\end{equation}
with the decrease of both terms as described above.
 
The second common feature of all models is that fluctuations in the
values of $\therPar$ tend to increase for decreasing $\Mh$, and this
is due to the ISM velocity field becoming more rich in substructure
for lighter DM haloes. Consistently with the hydrodynamical evolution,
the fluctuations in the \RD family are however smaller than in \CR and
\IS models, as the complexity of the ISM velocity field is
proportional to the amount of ordered rotation of the stellar
component. In particular, the large fluctuations of $\therPar$ in the
intermediate and light halo models of the \IS and \CR families are due
to the recurrent degassing events (with an increase of the velocity
components $\uR$ and $\uz$) in the equatorial plane, as
described in the previous Section.

An important difference between the three classes of models is instead
given by the average value of $\therPar$ that, for the \IS family,
is much lower then for the \CR and \RD families (for
which, for the reasons explained above, $\therPar$ can reach values
larger than unity). Instead, the \IS family is characterized by
$\therPar\simeq 0.1 - 0.3$, a remarkably lower degree of
thermalization. Being the large-scale kinematical support of the \CR and \IS 
families very similar, their significantly different
$\therPar$ must be explained by the
larger $\norm{\vphi\evphi -\uv}^2$ term that originates in the counter-rotating disc of the \CR family. The range of values of
$\therPar$ for the \IS family corresponds to the high-$\alpha$ models
in \citet{posacki.et.al2013}, and it provides part of the explanation for
the average lower X-ray luminosity of rotating models with respect to
non-rotating ones of the same mass (a feature that is found in
observed ETGs, see Sect. 1).
In Sect. 3.3 and 4 we will return to this point for some
additional considerations.

\subsection{$\Lx$, $\Tx$, and $\SigX$}
\begin{figure*}
\includegraphics[width=0.95\linewidth, keepaspectratio]{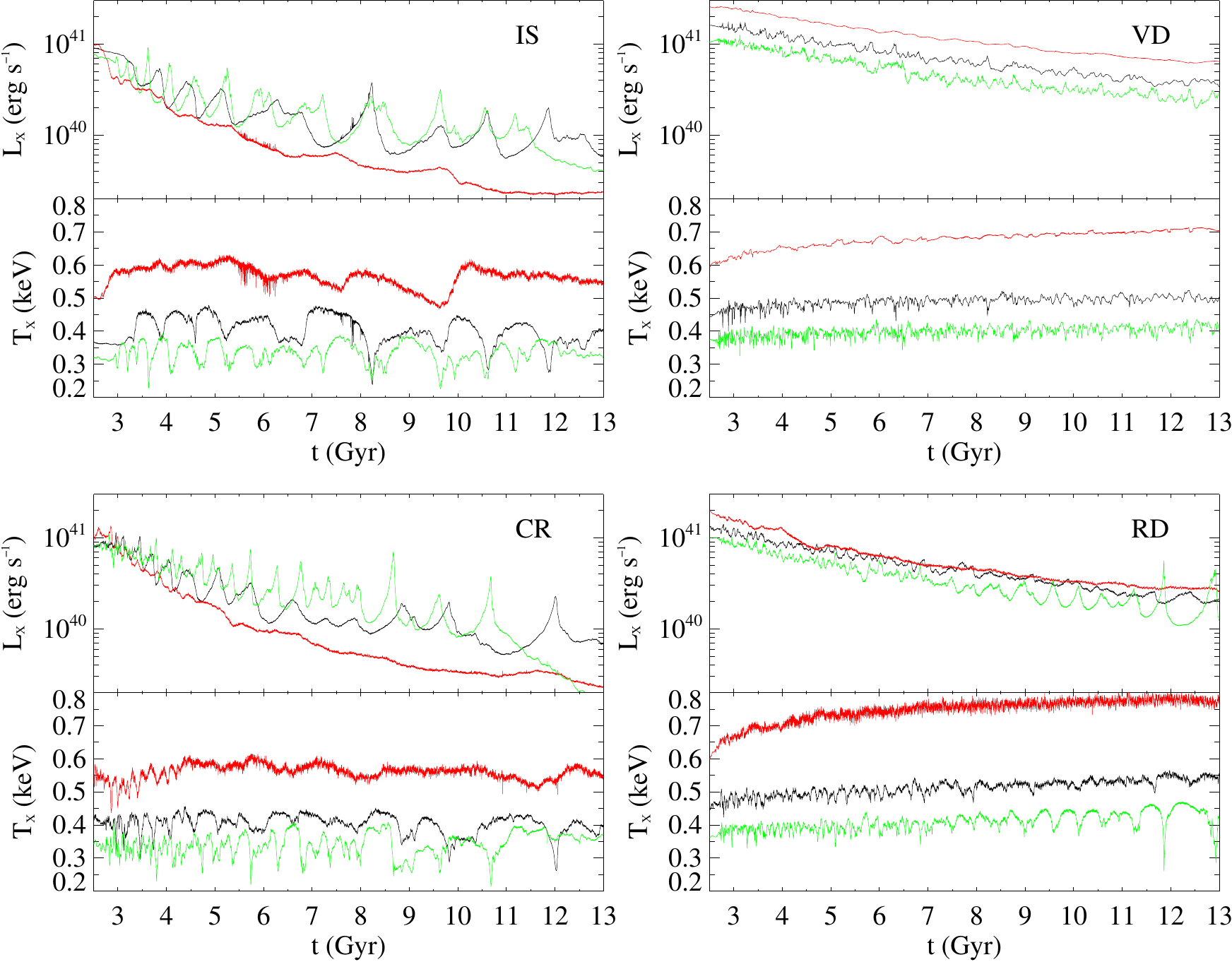}
\caption{Time evolution of $\Lx$ and X-ray emission weighted temperature
 $\Tx$ of the four families of models. Red, black, and green lines
 refer to the heavy, intermediate, and light DM haloes, respectively.}
\label{fig12}
\end{figure*}

The time evolution of the observationally important ISM diagnostics
$\Lx$ and $\Tx$ is shown for all models in Fig.~\ref{fig12}, and a list of
$\Lx$ and $\Tx$ values at the end of the simulations (13 Gyr) is given
in Tab.~\ref{tab:simulations}.

Important similarities and differences, due to the different internal
kinematical support and to the variable DM amount, are evident.
Concerning similarities, in all models the ISM X-ray luminosity
$\Lx$ decreases with time, broadly reflecting the decrease of the hot
gas content in the galaxies. Another similar behaviour is that, as
already discussed in Sect. 3.1, within each family the X-ray
luminosity weighted temperature $\Tx$ increases when increasing
the DM amount; in addition, $\Tx$ increases with time (as more
evident in the \VDh and \RDh models), due to the time
evolution of the specific heating of the injected material.

A major distinctive property of rotating models (\IS and \CR) with
respect to non-rotating ones (\VD and \RD) is instead the presence of
well defined oscillations in $\Lx$ and $\Tx$. These oscillations are
the result of the cyclic behaviour of the hydrodynamical evolution
typical of \IS and \CR models (Sects. 3.1.2 and 3.1.3). During each
oscillation (see for example that corresponding to the large peak in
$\Lx$ at around 8.2 Gyr for \ISi, mapped in Fig.~\ref{fig5}), $\Lx$ and $\Tx$
reach respectively a maximum and a minimum, all due to the onset of a
cooling phase. At the beginning of each cycle, the X-ray luminosity
is low, the galaxy is filled with the gas coming from stellar
evolution, and heated by the energy source terms. As the gas mass
rises, the cooling becomes more and more efficient due to the
compressional effect of the central outflow (Sect. 3.1.2), and $\Lx$
increases too.
This trend continues until a critical density is
reached, such that the radiative losses dominate over the heating sources, and the gas catastrophically cools; at this point
the peak in $\Lx$ is produced, with the associated sharp decrement in
$\Tx$. Finally, after the major radiative cooling phase has ended,
the hot gas density and $\Lx$ are low, and a new cycle starts. The
global pattern of an oscillation is always the same, and governed by
the mass injection and the radiative cooling rates. With time
increasing, oscillations become more distant in time, since the
refilling and the heating times become longer, due to the temporal
decay of both the mass injection rate and the number of SNIa events
(eqs. 16-17).

The second distinctive property of rotating models is that their $\Lx$
and $\Tx$ are always lower (for the same DM halo), than those of the
models that are non rotating on the large scale (the \VD and \RD
families).
This is an important feature, also for its observational implications
(Sect. 1), and thus deserves some consideration.
\textit{This
 important effect of rotation, as that of causing a cyclic behaviour of the flow, originates in a
 different flow evolution that in turn is due not just to the different energetic
 input, but also to the different angular momentum of the gas at large radii.}
In fact, a first explanation of the X-ray
under-luminosity and ``coolness'' of the \IS family, that seems natural, lies in the lack
of the $\sigmaphi$-term in their $\Lsigma$ (being $\sigmaphi$ replaced
by the ordered rotational field of the stellar component, see Fig.~\ref{fig:fig1}),
and in the result that $\therPar$ has low values ($<1$, to be inserted
in eq. 15).
However, this energy-based argument
does \textit{not} give the full explanation of the low $\Lx$ values; a
hint towards this conclusion is provided by the finding that $\Lx$ is
low also for the \CR family, with similar $\therPar$ values as for the \RD
one.
We established definitively that the different energy input to
the gas is not the sole explanation of the low $\Lx$ and $\Tx$ by performing
some ``ad hoc'' experiments in the \IS family. In practice,
while retaining their internal dynamical structure, we modified the
thermalization term in eq. (15), replacing it with the full
thermalization term of the \VD family.
Thus, from an hydrodynamical
point of view, the ISM of these models still rotates as dictated by
eq. (14), but its energy injection is equal to that of the \VD
family\footnote{In these tests the square brackets in eq. (15) was
 substituted with $\norm {\uv}^2 +\vphi^2 + \trace (\veldisp^2)$, so that
 from the virial theorem the sum of the last two terms equals $\trace
 (\veldisp^2)$ of \VD models. Actually the heating in these modified
 \IS models is even larger than in \VD ones, because the ISM velocity of the former contains also a relevant rotational component $\uphi$.}.
The results are interesting: on one side, $\Lx$ and $\Tx$ are higher
than in the \IS models; however, on the other side, $\Lx$ and $\Tx$ are
still characterized by large oscillations (typical of rotating models,
and absent in the \VD family), and they are still lower than in the
\VD family.
Having said this, one should also notice that, $\Lx$ differs more, by comparing \IS and \VD models, than the whole of $\Lrot$, which by itself shows that the energetic argument cannot account for the full gas behaviour (see
$\Lsigma$ and $\Lrot$ in Tab.~\ref{tab:simulations}).
Therefore, these experiments prove
that the X-ray under-luminosity and coolness of \IS and \CR models is not just due
to a reduction of the injection energy in them, but -more importantly- to the global evolution of the ISM induced by ordered rotation.

A few additional trends are shown by Fig.~\ref{fig12}. Time oscillations in
$\Lx$ and $\Tx$, for the rotating \IS and \CR models, become more and
more important for decreasing $\Mh$, reflecting the more structured
density, velocity and temperature fields of the ISM for lighter DM
haloes (see Sects. 3.1.2 and 3.1.3).
Another point is that the RD models always show the largest $\Tx$, for
any DM halo; this is due to their central, small, hot region
surrounding the small cold disc at their centres, a region that is not
present in the VD class. Finally, in the rotating families (\IS and
\CR), $\Lx$ is systematically lower for increasing DM halo, while the
opposite takes place for the two globally non-rotating families. This
trend is due to the global angular momentum stored at large radii in
\IS and \CR models, and its influence on the global behaviour of the
flow: in rotating models, the increase of $\Mh$ corresponds to an
increase of the total angular momentum, such that the galactic central
region (where most of the $\Lx$ comes from) is less dense of gas for
larger $\Mh$, due to the accretion of the gas on a larger cold disc.

We finally discuss the edge-on appearance of the X-ray surface
brightness maps $\SigX$ at the end of the simulation (for the intermediate
halo models; Fig.~\ref{fig13}). From the figure it is clear how large scale
galaxy rotation leads to flatter and ``boxy'' X-ray isophotes, and to
less concentrated X-ray emission, with respect to what is seen in
non-rotating VD and RD models. Also the ISM density
distribution is rounder in the VD, and more elongated along the
equatorial plane in the IS models.
In \IS models, this elongation and the consequent
``boxiness'' (already found in \citealt{brighenti.mathews.1996}, DC98) are due to the
rotational support on the equatorial plane, that prevents the gas from
flowing inward.
For what concerns a
comparison with the optical surface brightness distribution $\Sigma_\star$ of the parent galaxy (Fig.~13), in
non-rotating models $\SigX$ is rounder than $\Sigma_\star$ at
all radii, reflecting the rounder shape of the total isopotentials (mostly due to the DM).
On the other side, rotation proves to be effective in
determining a significantly flatter $\SigX$, that becomes more similar
to $\Sigma_\star$, especially in the inner galactic region (within $\simeq
10$ kpc); however $\SigX$ is never flatter than $\Sigma_\star$.
$\SigX$ becomes more spherical at large radii, and the boxiness decreases with
radius, but it is still present at $R\simeq 30$ kpc.

\subsection{Comparison with observed X-ray properties}
\begin{figure*}
\includegraphics[width=0.9\linewidth, keepaspectratio]{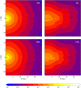}
\caption{Edge-on 0.3--8 keV surface brightness of the ISM at 13 Gyr,
 for the intermediate halo models \VDi, \ISi, \RDi and \CRi; the
 brightness values on the colour-bar are given in erg s$^{-1}$
 cm$^{-2}$. Superimposed are the isophotes ($\Sigma_\star$) obtained by projecting the
 galaxy stellar density distribution, starting from $10^4$
 $M_{\odot}$ pc$^{-2}$ on the innermost contour, and decreasing by a
 factor of ten on each subsequent contour going outwards.}
\label{fig13}
\end{figure*}

As a general check of the reliability of the gas behaviour obtained
from the simulations, we consider here a broad comparison with the
observed X-ray properties of the Sombrero galaxy (whose structure was
taken as reference), and of flat ETGs. In the Sombrero galaxy,
diffuse hot gas has been detected in and around the bulge
region with \textit{XMM-Newton} and \textit{Chandra} observations \citep{li.et.al2007, li.et.al.2011}, extending to at
least $\simeq 23$ kpc from the galactic centre, roughly as obtained here (Fig.~\ref{fig13}).
The X-ray emission is stronger along the major axis than along the minor axis, and can be
characterized by an optically thin thermal plasma with $kT\simeq 0.6$
keV, varying little with radius.
The total 0.3--2 keV luminosity is
$\Lx=2\times 10^{39}$ erg s$^{-1}$, and the hot gas mass is $\simeq
5\times 10^8 M_{\odot}$ \citep{li.et.al.2011}.
The gas has a supersolar metal abundance, not expected for accreted intergalactic
medium, thus it must be mostly of internal origin, as that studied
in this work.
In a simple spherical model for the hot gas, originating from internal mass sources heated by SNIa's, a supersonic galactic wind develops for a galaxy potential as plausible for Sombrero, with an
$\Lx$ far lower than observed \citep{li.et.al.2011}.
A flow different from a wind, and as found by our 2D hydrodynamical
simulations, may provide the correct interpretation of the observed X-ray properties.
Among the suite of models run
in our work, an $\Lx$ value comparable to that observed is reached at
the present epoch by the \IS models (Tab.~\ref{tab:simulations}); further, the $\Tx$ of
the \ISh model is very close to the observed one, while \ISi and \ISl
have a lower $\Tx$ (note though that the $\Lx$ and $\Tx$ values
in Tab.~\ref{tab:simulations} refer to the whole computational grid, corresponding to a
physical region larger than that used for the X-ray observations).

Our work thus shows how it is crucial to account for the proper
shape of the mass distribution (e.g., bulge, disc and dark matter
halo), as well as for the angular momentum of the mass-losing stars,
to reproduce the hot gas observed properties. For
example, \VD-like models predict $\Lx$ larger by an order of magnitude, and
inconsistent with those of Sombrero. Another feature clearly
requiring angular momentum of the stars is provided by the observed
X-ray isophotes, that show a boxy morphology in the inner regions \citep{li.et.al.2011}, as obtained by our models only in case of
rotation.

Finally, note that the hot gas emission in Sombrero is lower than
predicted by the best fit $\Lx-L_{\rm K}$ correlation observed for
ETGs \citep[e.g.][]{boroson.et.al2011}, as shown by \citet{pellegrini1999, pellegrini2005, pellegrini2012}. The X-ray luminosity could be reduced in Sombrero by the
effects of rotation, as explained in Sect. 3.3.

Moving to X-ray observations of S0 galaxies, a
\textit{Chandra} survey of their X-ray properties has been recently performed by \citet{li.et.al.2011}.
They tend to have significantly lower
$\Lx$ than elliptical galaxies of the same stellar mass.
While \citet{li.et.al.2011} focussed on the possible cold-hot gas interaction to
find an explanation \citep[see also ][]{pellegrini.et.al2012}, we can suggest
that rotation could have an important effect.
A case S0 study is
NGC5866 \citep{li.wang.li.chen2009}, where the morphology of the hot gas
emission appears rounder and more extended than that of the stars (the
galaxy is seen edge-on), and again the X-ray isophotes have a boxy
appearance in the inner region \citep[see Fig. 1b in][]{li.wang.li.chen2009}.
However, the stellar mass is much lower than for Sombrero ($\Mstar \simeq
3\times 10^{10}\Msun$), so we cannot directly compare $\Lx$ and $\Tx$
of our modelling with the observed hot gas properties of NGC5866 (the
latter is just $7\times 10^{38}$ erg s$^{-1}$).


\section{Discussion and conclusions}\label{conclusions}
In this work we studied the effects of the stellar kinematics on the
ISM evolution of flat ETGs, focussing in particular on a
representative S0 galaxy model, tailored on the Sombrero galaxy. We
considered four different families of galaxies, characterized by the
same stellar distribution, and by a spherical DM halo of fixed
scale-lenght, with three different total mass values. In the first
family (\IS), the galaxy flattening is entirely supported by ordered
rotation of the stellar component, while the velocity dispersion tensor is everywhere
isotropic. In the second family (\VD), the galaxy flattening is all
due to stellar azimuthal velocity dispersion, i.e, the models are fully
velocity-dispersion supported. The other two families are variants of
the first two: in the \CR family a thin counter-rotating stellar disc
is placed in the central region of \IS models; in the \RD family, a
rotating thin stellar disc is placed at the centre of the
non-rotating \VD models. The standard sources of mass and energy for the ISM are
adopted, while we neglect star formation (and more in general diffuse
mass sinks due to local thermal instabilities), and feedback effects
due to a central supermassive black hole. The simulations have been
performed in cylindrical symmetry, and cover the ISM evolution for 11 Gyr.
The main results can be summarized as follows.

In all models, the ISM velocity and density at early times are
symmetric with respect to the equatorial galactic plane, but soon this
reflection symmetry is lost, and the hydrodynamical evolution of
rotating models becomes more complicated than that of non-rotating
ones. In \VD and \RD families the cooling gas tends to flow directly to
the galaxy centre, while conservation of angular momentum leads to the
formation of a cold rotating gaseous disc in IS ones. Moreover, the
flow in rotating models is spatially decoupled (as already found in
lower-resolution simulations, DC98), and shows large-scale meridional
circulation outside a $V$-shaped region containing the galaxy
equatorial plane, while inside this region the ISM is radially
supported against the gravitational field by azimuthal rotation. 
These features are confirmed by a few runs with a number of
gridpoints increased by a factor of four, reaching a spatial resolution
of 15 pc in the central regions. In
general, with increasing time the ISM velocity tends to
decrease. The gas is not outflowing from the galactic outskirts in a
significantly larger amount in rotating than in non-rotating families,
as shown by the similar $\Mesc$ values in Tab.~\ref{tab:simulations}. Indeed, the major
effect of ordered rotation in our models (that refer to a massive galaxy)
is not that of making the gas
less bound, but that of changing the behaviour of the gas.

A remarkable difference between globally rotating (\IS and \CR) and
globally non-rotating (\VD and \RD) models is represented by large
secular oscillations of $\Lx$ and $\Tx$ in the former (see also DC98,
\citealt{negri.et.al.2013proceeding}). These oscillations are due to
periodic cooling episodes in the $V$-shaped region of the rotating families,
accompanied with degassing events along their equatorial plane
(however limited to within few tens of kpc). Furthermore, in \IS
models a central hot region of radius comparable to that of the disc
develops above and below the disc, due to the lower gas density there
with respect to \VD models. Outside of this hot sphere, the \IS
temperature is lower than that of \VD models.

The X-ray luminosity $\Lx$ is largest for the velocity dispersion
supported \VD and \RD families, and the highest X-ray emission-weighted
temperatures are shown by the \RD family, at any given DM halo mass.
The strong rotators (\IS and \CR) are characterized by $\Lx$ and $\Tx$
significantly lower than for non-rotating models; for example, $\Lx$
can be more than a factor of ten lower, at the same DM halo mass (see
Tab.~\ref{tab:simulations}). In all cases, $\Lx$ and $\Tx$ are within the range of
observed values for galaxies of similar optical luminosity and central
velocity dispersion. \citet{sarzi.et.al.2013}, following CP96,
suggested that the different $\Lx$ of
slow and fast rotators could be due to fast rotators, being on average flatter, 
being also more prone to loose their hot gas; here we find that the different 
$\Lx$ is indeed due to a lower hot gas content of rotating systems, but this 
is not produced by a larger fraction of escaped gas, but instead by a larger
amount of hot gas that has cooled below X-ray temperatures. 
These results are confirmed also by 
an ongoing investigation of the flow properties for a large set of galaxies
with different shapes and internal kinematics (Negri et al., in preparation).
In agreement with CP96 we also find, though, that in low mass galaxies, generally tending to develop outflows,
rotation favours gas escape.

For increasing DM halo mass, the ISM velocity fields become more
regular, with less substructure, and $\Tx$ increases, while $\Lx$
behaves differently: in rotating models, $\Lx$ decreases with
increasing $\Mh$, while it increases in non-rotating models. Rotating
\IS and \CR models with light DM haloes at late times develop a polar
wind.

The (edge-on) X-ray isophotes are rounder than the stellar isophotes,
as expected due to the round shape of the total gravitational
potential. However, in rotating models the X-ray isophotes tend to be
boxy in the inner regions, while in non-rotating models they are almost
spherical.

Note finally how the gas evolution and overall properties of the \IS and \CR 
families on one side, and of the \VD and \RD on the other, are remarkably similar, 
i.e., the presence of centrally rotating stellar discs does not alter
significantly the global flow behaviour.

In order to quantify the amount of galactic ordered rotation which is
actually thermalized, we computed the thermalization parameter $\therPar=
\Lstr/\Lrot$, that is the ratio of the heating due to difference between the
streaming velocity of the stars and the ISM velocity, and the heating that would be
provided by the stellar streaming if stars were moving in an ISM at rest.
We found that $\therPar$ is substantially less than unity in \IS models, while it is of the order of unity
or more in \CR and \RD ones; $\therPar$ increases for lower DM contents.
This shows that the different behaviour of \VD and \IS families
is not entirely due to the different amount of thermalization of the
stellar motions, but rather to the impact of angular momentum on the flow at
large scales. In fact, despite of their different $\therPar$,
all the main features of the \IS family 
are still present in the \CR one, including the well defined
oscillations in $\Lx$ and $\Tx$ that are absent in \VD and \RD
models. 

The parameter $\therPar$ is used in works involving the global energy
balance of the gas \citep[e.g., CP96;][]{posacki.et.al2013}.
\citet{posacki.et.al2013} showed that low values of $\therPar$ go in the direction of accounting for the relatively low
values of $\Lx$ and $\Tx$ in flat (and rotating) galaxies when
compared to the values of their non-rotating counterparts.
Also \citet{sarzi.et.al.2013} suggested that the kinetic energy
associated with the stellar ordered motions may be thermalized less
efficiently to explain why fast rotators seem confined
to lower $\Tx$ than slow rotators.
The fact that $\therPar$ is substantially less than
unity in \IS models could provide support to an energetic
interpretation of the X-ray under-luminosity of flat and rotating
galaxies, when compared to non-rotating ones of similar optical
luminosity. The results for \CR models, however, show that the lack of
thermalization of ordered rotation cannot be the {\it only}
explanation of the low values of $\Lx$: for these models $\therPar$ is
in fact of the order of unity, yet their $\Lx$ is similar to that of
the corresponding \IS models, in which $\therPar\simeq 0.1-0.2$. Thus,
even if the presence of a stellar counter-rotating thin disc can
increase the thermalization of the ordered motions from 10-20 per cent
in the isotropic rotators, up to 100 per cent or more, {\it additional
 phenomena related to the global angular momentum (mainly stored at
 large radii) influence the behaviour of the flow}, and produce a
difference in $\Lx$ and $\Tx$. An additional glaring evidence for the
important role of angular momentum is provided by the fact that, in
rotating models (both \IS and \CR), $\Lx$ {\it decreases} for
increasing $\Mh$, that is associated with an {\it increase} of
galactic rotation. Summarizing the key points of this work, $\Lx$ is
significantly decreased not due to a larger degassing, or to a lower
energetic input (due to a ``missing'' part in $\Lkin$), but by crucial
angular momentum-related effects; pure energetic arguments cannot
fully account for the changes in the overall gas properties (e.g.,
$\Lx$, $\Tx$), and thus cannot solve the problem of the X-ray
under-luminosity, and ``coolness'', of rotating galaxies. While a
whole exploration of the flattening vs. rotation roles is deferred to
a subsequent work, here we can comment on the expected trend
with flattening. A decrease of galaxy flattening will be accompanied by
a decrease of the importance of $v_{\varphi}$, according to the Jeans
equations; and, in the limit of null flattening, two-integral Satoh models
become fully isotropic, spherical systems. Thus, for rounder shapes, all the
effects connected with stellar streaming will necessarily decrease. 

The present study can be relevant to the topic of black hole fuelling.
One of the most debated aspects of SMBH accretion is how gas is carried to the centre of galaxies, 
especially in presence of rotation; another aspect is whether the source of fuel is a
hot, roughly spherical atmosphere, from which accretion is almost steady, or it lies in cold material
that sporadically and chaotically accretes \citep[e.g.,][]{novak.etal2013, werner.etal2013, russell.etal2013}. 
Related to these important issues, the present investigation shows that in velocity dispersion supported systems
accretion is more hot and radial, with a large fraction of the total input from 
stellar mass losses flowing straight to the centre; 
in systems more supported by rotation, instead, the central density of hot ISM is lower, 
the mass accreted towards the centre is very small, 
and a cold rotating disc provides a large reservoir of cold gas, 
that can lead occasionally to clumpy multiphase accretion.
Moreover, as anticipated in the Introduction, 
the presence of a counter-rotating structure affects the central feeding: 
the simulations show that, by reducing the amount of local angular 
momentum, accretion in the central grid is favoured with respect to what happens 
in pure isotropic rotators.

\section*{Acknowledgements}
We thank the anonymous referee for comments that improved the presentation, 
James Stone for useful advices on the code, and Silvia
Posacki for providing the galaxy models. L.C. and S.P. are supported
by the Italian grants Prin MIUR 2008 and the Prin MIUR 2010-2011,
project `The Chemical and Dynamical Evolution of the Milky Way and
Local Group Galaxies', prot. 2010LY5N2T. This material is based upon
work supported in part by the National Science Foundation under Grant
No.~1066293 and the hospitality of the Aspen Center of Physics.
\balance 
\bibliographystyle{mn2e}
\bibliography{citations}
\label{lastpage}

\end{document}